\documentclass[superscriptaddress,aps,pre,twocolumn,floatfix]{revtex4-1}

\usepackage{amsmath}
\usepackage{amssymb}
\usepackage{refcount}
\usepackage{siunitx}
\usepackage{graphicx}
\usepackage{hyperref}
\usepackage{braket}
 \usepackage[capitalise]{cleveref}
\usepackage{microtype}
\usepackage{bm}
\usepackage[english]{babel}
\newcommand{\abs}[1]{\left| #1 \right|}

\newcommand{\br}{\mathbf{r}}

\newcommand{\bp}{\mathbf{p}}

\newcommand{\comm}[2]{\left[ #1 , #2 \right]}

\usepackage[cmyk]{xcolor}

\begin{document}
\author{Gert Brodin}
\email{gert.brodin@umu.se}
\author{Haidar Al-Naseri}
\email{haidar.al-naseri@umu.se}
\author{Jens Zamanian}
\author{Greger Torgrimsson}

\affiliation{Department of Physics, Ume{\aa} University, SE--901 87 Ume{\aa}, Sweden}
\author{Bengt Eliasson}

\affiliation{SUPA, Department of Physics, University of Strathclyde, Glasgow G4 0NG, UK}
\title{Plasma dynamics at the Schwinger limit and beyond}

\begin{abstract}
Strong field physics close to or above the Schwinger limit are typically studied with vacuum as initial condition, or by considering test particle dynamics. However, with a plasma present initially, quantum relativistic mechanisms such as Schwinger pair-creation are complemented by classical plasma nonlinearities. In this work we use the Dirac-Heisenberg-Wigner formalism to study the interplay between classical and quantum mechanical mechanisms for ultra-strong electric fields. In particular, the effects of initial density and temperature on the plasma oscillation dynamics are determined.    
\end{abstract}

\maketitle
\section{Introduction }
In recent years, much research has been devoted to strong field physics where various QED effects come into play, see e.g. Refs. \cite{review 1, recent review} for review articles. A key motivating factor has been the rapid evolution of laser intensities \cite{review 1, recent review, ELI}. Nevertheless, in spite of the steady progress, peak electric fields for current lasers are still well below the critical field strength $E_{cr}=m^2c^3/e\hbar\approx 1.3\times10^{18}{\rm V/m}$ \cite{Schwinger}. It should be noted, however, that there are proposals trying to use plasma based schemes in order to magnify present day laser fields to reach the Schwinger limit \cite{plasma mirror}. 

Research on ultra-high intensities, i.e. close to or above the Schwinger limit, has been particularly concerned with Schwinger pair-creation in vacuum \cite{Gies,Gies2,optimal pulse,Aleksandrov:2019ddt,Kohlfurst:2019mag}. Problems of study include e.g., understanding the interplay between temporal and spatial variations of the fields \cite{Gies2,Aleksandrov:2019ddt,Kohlfurst:2019mag}, comparing different computational schemes \cite{Gies,Aleksandrov:2019ddt}, or optimizing the geometry of colliding laser pulses to maximize pair-creation \cite{optimal pulse}. 

For sufficiently strong fields, large number of pairs are created,  leading to plasma currents modifying the original fields, such that the back-reaction through Ampère's law is relevant \cite{Cooper-89,Kluger,Ruffini, Bloch,Hebenstreit:2013qxa,Buyens:2016hhu,Kasper:2014uaa,Kasper:2016mzj,Kasper:2015cca,Gelis:2013oca,Gelis-Tanji-review,Tanji:2008ku,Kluger,Kampfer,Gold:2020qzr}. Starting the dynamics with a Sauter-pulse \cite{Sauter} or a constant initial electric field, the system will undergo plasma oscillations, with a plasma frequency proportional to the peak electric field.  

In the present paper we intend to study how the dynamics is modified when a plasma of high density is present initially. In this case several new questions enter the picture. How does the dynamics depend on the plasma temperature and density? What is the interplay between classical plasma nonlinearities, and those due to quantum relativistic effects? To what extent can pair creation be blocked if the low-energy states are already occupied? 

For this purpose, we will use the Dirac-Heisenberg-Wigner (DHW) formalism \cite{Birula} in order to address the above questions numerically. 
After theoretical preliminaries, including settling the issue of charge renormalization, we study plasma oscillations in an electron-ion plasma focusing on the homogeneous limit.  
It is worth noting that with a plasma present from the start, the energy density necessary for pair-creation can equally well be stored in the plasma current. In this case the initial electric field is zero, but the plasma kinetic energy is converted into an electric field of sufficient magnitude.

No restrictions on the electric field amplitude will be made, and thus the dynamics will be studied also for field strengths well beyond the critical field. For a field strength of the order of the critical field or higher, the plasma oscillations increase their frequency due to the pair-creation process, at the same time as they are damped.

This paper is organized as follows: 
In \cref{DHW-formalism} we briefly discuss the derivation of the DHW-formalism and present a reduced system applicable for electrostatic fields. 
In \cref{Renormalization} we present the calculations relevant to charge renormalization. Next, the numerical results are presented in \cref{Numerical results}. Here we show the scaling of the damping with the initial electric field amplitude and with the plasma parameters, study issues related to Pauli blocking, and relate our findings to previous results starting from vacuum. Finally, we discuss potential competing mechanisms in  \cref{Competing mechanisms} and draw the final conclusions in \cref{Conclusion}.   

\section{The DHW-formalism:} 
\label{DHW-formalism}

The analysis will be based on the DHW-formalism, first presented in Ref.
\cite{Birula}. Slightly different derivations have also been presented by Refs.
\cite{Vasak-87,Gies,Sheng}. As the presentation given in these works are fully satisfactory, we
will not give details of the derivations here, just point out the main
features:

\begin{enumerate}
\item The derivations is based on a Gauge invariant Wigner transformation of
the Dirac equation, producing a 4 by 4 Wigner matrix (due to the Dirac four
spinors), where $\hat{W}(\mathbf{r},\mathbf{p},t)$  is depending on the phase
space variables
\begin{multline} \label{wigtrans}
\hat{W}(\br,\bp,t) = 
\\
-\frac{1}{2}\int d^3z
\exp \bigg(-i\textbf{p}\cdot\textbf{z} -ie\int^{1/2}_{-1/2} d\lambda \mathbf{z}\cdot \mathbf{A}(   \br+\lambda\mathbf{z},t )\bigg) 
\\ 
\times \comm{ \hat{\Psi}(\br+\mathbf{z} /2,t )} { \hat{\bar{\Psi}} (\br -\mathbf{z}/2}.,
\end{multline}
Note that the Wilson-line factor has been used to ensures the gauge-invariance. The Wigner function $W(\br, \bp,t)$ is defined as the expectation value of the Wigner operator
\begin{equation}
\label{Wigner_function}
    W(\br,\bp,t)=\braket{\Omega |\hat{W}(\br,\bp,t)|\Omega}. 
\end{equation}

\item In the derivation, the vacuum fluctuations are dropped in comparison
with the electromagnetic mean field, which is a particularly accurate
approximation for strong field problems.

\item The 16 components of $W(\mathbf{r},\mathbf{p},t)$ can be split using
various projections, to formulate the theory in a physically attractive way.
We have used the standard split of Refs. \cite{Birula, Gies}, which divides $W(\mathbf{r},%
\mathbf{p},t)$ into phase space functions for quantities like, mass, charge
density, current density, spin, and magnetization; see Ref. \cite{Birula} for the full
details. 

\item Since the DHW-functions couple to current and charge sources, the
system is closed by Maxwell's equations.
\end{enumerate}
With the closed system in place, further simplifications can be made
depending on the geometry of the electromagnetic field. For the case of a
one-dimensional electrostatic field \cite{Estat-note} $\mathbf{E}=E(z,t)\mathbf{\hat{z}}$, the
full system (see Ref. \cite{Perpendicular}) can be reduced to 

\begin{align}
\label{PDE_System}
    D_t\chi_1(z,\bp,t)&= 2\epsilon_{\bot}(p_{\bot}) \chi_3(z,\bp,t)- \frac{\partial \chi_4}{\partial z} (z,\bp,t)\notag\\
    D_t\chi_2(z,\bp,t) &= -2p_z\chi_3(z,\bp,t)\\
    D_t\chi_3(z,\bp,t)&= -2\epsilon_{\bot}(p_{\bot}) \chi_1(z,\bp,t) +2p_z\chi_2(z,\bp,t)\notag\\
    D_t\chi_4(z,\bp,t)&= -\frac{\partial \chi_1}{\partial z}(z,\bp,t)\notag 
\end{align}
together with the Ampère's law
\begin{equation}
\label{Ampers_law}
\frac{\partial E}{\partial t}=-\frac{e}{(2\pi)^3}\int \chi_1 d^3p
\end{equation}
where $D_t=\partial/\partial t +eE\partial/\partial p_z$ and $\epsilon_{\perp}=\sqrt{m^2+p_{\perp}^2}$, and where $p_{\perp}$ is the perpendicular momentum. Apart from constants (see Ref. \cite{Birula,Perpendicular}), the variable $\chi_{1}$ is the
(phase space) current density,  $\chi_{2}$ mass density, $\chi_{3}$ spin density, and $\chi_{4}$ charge density. \ 

First we rewrite the system \cref{PDE_System} to
 simplify the numerical calculations. In this context it is important to note that $\chi_1$ and $\chi_2$ have nonzero vacuum contributions, associated with the expectation values of the free Dirac field operators, (see e.g. Ref. \cite{Birula,plasma quantum kinetic review}) where the vacuum expressions are given by
  \begin{align}
 \chi_{1vac}&=-\frac{2 p_z}{\epsilon}\notag \\ \chi_{2vac}&=-\frac{2\epsilon_{\perp}}{\epsilon}.
 \end{align}
 Here $\epsilon=\sqrt{m^2+p^2}$ , and we introduce new variables $\Tilde{\chi}_i(z,\bp,t)$ as the deviation from the vacuum state, i.e. we let
 \begin{equation}
     \Tilde{\chi}_i(z,\bp,t)=\chi_i(z,\bp,t)- \chi_{i {\rm vac}}(\bp).
    \end{equation}
    Note that $\Tilde{\chi}_{3,4}={\chi}_{3,4}$. 
 Secondly, we switch to the canonical momentum. Using the Weyl gauge where the scalar potential is zero such that $E=-\partial A/\partial t$, the $z$-component of the kinetic momentum is replaced by $q=p_z+eA$. With $q$ as one of the independent variables, the operator $D_t$ simplifies to $D_t=\partial/\partial t$. Our third modification is to switch to dimensionless variables.  The normalized variables are given by, $t_n= \omega_c t$, $q_n=q/mc$, $p_{n\perp}=p_{\perp}/mc$, $E_n=E/E_{cr}$, $A_n=eA/mc$, where $\omega_c =(mc^2/\hbar)$ is the Compton frequency, and we note that the DHW-functions are already normalized. For notational convenience, we omit the index $n$ in what follows. Finally, we consider the homogeneous limit,  making the variable $\Tilde{\chi}_{4}=0$. The equations to be solved numerically now read: 
\begin{align}
\label{PDE_System4}
    \frac{\partial \Tilde{ \chi}_1}{\partial t} (q,p_{\bot},t)&= 2\epsilon_{\bot} \Tilde{\chi}_3 + 2E\frac{\epsilon_{\bot}^2}{\epsilon^3} \notag\\
      \frac{\partial  \Tilde{\chi}_2}{\partial t} (q,p_{\bot},t )&= -2(q-A)\Tilde{\chi}_3-2(q-A)E\frac{\epsilon_{\bot}}{\epsilon^3}\notag\\
      \frac{\partial \Tilde{\chi}_3}{\partial t} (q,p_{\bot},t)  &=
      -2\epsilon_{\bot} \Tilde{\chi}_1 
      +2(q-A)\Tilde{\chi}_2 
      \end{align}
     
      with Ampère's law
      \begin{equation}
\label{Ampers_law2}
\frac{\partial E}{\partial t}=- \eta \int \chi_1 d^2p
\end{equation}
where the dimensionless factor $\eta=\alpha/\pi\approx 2.322 \times10^{-3}$. 
Note that due to cylindrical symmetry the azimuthal integration has already been carried out and hence $d^2p=p_{\bot}dqdp_{\bot}$.
\section{Charge renormalization}
\label{Renormalization}
Generally, the momentum integrals in Ampère's law in \cref{Ampers_law2} are subject to
UV-divergences, i.e. the vacuum contribution to the integrand scales as $%
1/\varepsilon $ for large energy (momentum) leading to a logarithmic divergence. For a numerical solution this
could possibly be solved by a numerical cut-off acting as an effective
regularization. To assure the soundness of the numerical scheme, however, we
must analyze this issue in more detail. From \cref{PDE_System4} we immediately obtain 
\begin{equation}
\frac{\partial ^{2}\tilde{\chi}_{3}}{\partial t^{2}}+4\varepsilon ^{2}\tilde{%
\chi}_{3}=-\frac{4\varepsilon _{\bot }E}{\varepsilon }+2E\tilde{\chi}_{2}.
\label{chi3}
\end{equation}%
For a modest plasma density, with a plama frequency well below the
pair-creation resonance (i.e. $\partial _{t}\ll 2\varepsilon $), the left
hand side operator of Eq. (\ref{chi3}) can be inverted to all orders in an expansion $\,$in the
small operator $(1/\varepsilon )\partial _{t}$. Formally solving Eq. (\ref%
{chi3}) for $\tilde{\chi}_{3}\,\ $we get%
\begin{equation}
\tilde{\chi}_{3}=-D^{-1}\left[ \frac{4\varepsilon _{\bot }E}{\varepsilon }-2E%
\tilde{\chi}_{2}\right]   \label{chi3-solved}
\end{equation}%
where the inverse operator is 
\begin{equation}
D^{-1}=\left[] 1-\frac{1}{4\varepsilon ^{2}}\frac{\partial ^{2}}{\partial
t^{2}}+ \left( \frac{1}{4\varepsilon ^{2}}\frac{\partial ^{2}}{\partial t^{2}} \right)^2%
\right] \frac{1}{4\varepsilon ^{2}}  \label{InverseD}
\end{equation}%
Keeping only terms up to $\partial ^{2}/\partial t^{2}$, we use 
Eqs. (\ref {chi3-solved}-\ref{InverseD}) in \cref{PDE_System4} to get an equation for the phase space
current $\tilde{\chi}_{1}$, which reads%
\begin{equation}
\frac{\partial \tilde{\chi}_{1}}{\partial t}=2\frac{\varepsilon _{\bot }^{2}}{%
\varepsilon ^{2}}\frac{\partial ^{2}}{\partial t^{2}}\left( \frac{E}{%
4\varepsilon ^{3}}\right) +4\varepsilon _{\bot }\left[ \left( 1-\frac{1}{%
4\varepsilon ^{2}}\frac{\partial ^{2}}{\partial t^{2}}\right) \frac{E\chi
_{2}}{4\varepsilon ^{2}}\right] .  \label{chi1-derivative}
\end{equation}%
Finally, combining \cref{chi1-derivative} with the time derivative of
Ampère's law (\ref{Ampers_law2}), we obtain.
\begin{multline}
\frac{\partial ^{2}E}{\partial t^{2}}\left( \frac{1}{\eta_B}+ 2\int \frac{\varepsilon
_{\bot }^{2}}{\varepsilon ^{5}}d^{3}p\right) =\\
- \int 2\frac{\varepsilon
_{\bot }^{2}}{\varepsilon ^{2}}\left[ 2\frac{\partial E}{\partial t}\frac{%
\partial }{\partial t}\left( \frac{1}{\varepsilon ^{3}}\right) +E\frac{%
\partial ^{2}}{\partial t^{2}}\left( \frac{1}{\varepsilon ^{3}}\right) %
\right] d^{3}p- \frac{\partial j_{p}}{\partial t}  \label{Amp-renorm}
\end{multline}%
where we have renamed $\eta=\eta_B$ to highlight that this corresponds to the bare value of the coefficient, obtained by using the bare value of the charge $e_B$, and noting that $ \eta_B\propto e_B^2  $. Here we have introduced the plasma current  
\begin{equation}
j_{p}=\int 4\varepsilon _{\bot }\left[ \left( 1-\frac{1}{4\varepsilon ^{2}}%
\frac{\partial ^{2}}{\partial t^{2}}\right) \frac{E\tilde{\chi}_{2}}{%
4\varepsilon ^{2}}\right] d^{3}p  \label{plasma-current}
\end{equation}%
which is the part of the current that is proportional to the mass density.
Recall that $\tilde{\chi}_{2}$ is defined as the deviation from the
vacuum state, and thus new particles that are created will also contribute
to the plasma current. However, our main concern here is the vacuum contribution of the left hand side of \cref{Amp-renorm}, containing the 
the integral $%
2\int (\varepsilon _{\bot }^{2}/\varepsilon ^{5})d^{3}p$ which has a logarithmic
divergence. As deduced in Refs. \cite{Bloch,linear}, by introducing a
regularization (i.e. a cut-off $\Lambda $ in momentum space) and replacing the
bare value of the electron charge $e_B$ with the physically measurable one, $e_R$, according to
\begin{equation}
   e_R^2 = \frac{e_B^2} {1+\frac{e_B^2}{24\pi^2 } \ln{\Big(\frac{\lambda}{m}\Big)}} 
\end{equation}
Ampère's law gets its standard form. Since $ \eta_B\propto e_B^2  $, this  corresponds to
\begin{equation}
\frac{1}{\eta_B}+2 \int^{\Lambda }\frac{%
\varepsilon _{\bot }^{2}}{\varepsilon ^{5}}d^{3}p= \frac{1}{\eta_B}+
\frac{\ln \Lambda  }{6} \rightarrow \frac{1}{\eta_R}
\label{renorm-factor}
\end{equation}%
The remaining part of the vacuum current,
written on the right hand side of \cref{Amp-renorm}, is nonzero but finite, due to the temporal variations of $\varepsilon
(t)=m\left( 1+p_{\bot }^{2}+(q-A(t))^{2}\right) ^{1/2}$. Moreover, as $1/\eta_R\gg1$, the relative importance of the non-vanishing vacuum contribution, also known as vacuum polarization, is small. 

From a theoretical point of view, when the back-reaction in Ampère's law is
included, it is important to confirm that the UV-divergences can be handled by a renormalization. However, for our case to be studied, typically with a numerical cut-off of the order $%
\Lambda \sim 70$, and with $\eta =\alpha /2\pi ^{2}$, the slow logarithmic divergence term gives a finite contribution that
in practice is very small, of the order \thinspace $0.002$ of the
plasma current or smaller. Accordingly, for the parameters used here, the difference between the bare charge and the measurable one is negligible. It should be stressed, though,  that this
conclusion is largely dependent on technical details. For a much larger
numerical cut-off, or for an enhanced value of the fine structure constant
(as is sometimes used to speed up numerical schemes, see e.g. Ref. \cite{Bloch}), it could
very well be necessary to use the renormalized expression. The above
calculation shows that it is straightforward to do so also when applying the
DHW-formalism. However, due to the smallness of this correction for the parameters of our case, we have not done this in the numerical calculations that follow.

\section{Numerical results}
\label{Numerical results}
\subsection{Preliminaries}
Equations (\ref{PDE_System4})-(\ref{Ampers_law2}) are solved numerically using a phase corrected staggered leapfrog method. While this is straightforward in principle, the full problem with three independent variables $\chi=\chi(q,p_{\perp},t)$ is still numerically demanding when run on a standard workstation. One reason is the strongly relativistic motion (gamma factors sometimes exceeding 50) requiring a high value of the momentum cut-off in $q$-space. 

That the numerical scheme produce sound results is confirmed by studying the energy conservation law. An energy conservation law of the system (\ref{PDE_System4})-(\ref{Ampers_law2})  can be written in the form 
\begin{equation}
    \frac{d}{dt}\left[ \frac{E^2}{2}+\eta\int[\tilde\chi_2+(q-A)\tilde\chi_1] d^2p \right]=0
\end{equation}
Due to the back-reaction in Ampère's law, containing both the conduction current and polarization current \cite{Kluger,Bloch}, the energy density of the electric field is diminished at the same rate as the kinetic energy of the initial plasma and of the produced pairs grow. For the numerical resolutions used in the runs presented below, the total energy of the system is conserved  within a relative error typically less than $10^{-4}$. Note that the total energy includes also the growing rest mass energy due to particle creation.  

As initial conditions in the runs we let $E(t=0)=E_0$ and $A(t=0)=0$. Moreover, the normalized background variables for $t=0$ are taken as 
\begin{align}
    \tilde{\chi}_1&=\frac{2q f(q,p_{\bot } ) }{(1+q^2+p_{\perp}^2)^{1/2}}\notag\\ 
    \Tilde{ \chi}_2&=\frac{2f(q,p_{\bot } ) )}{(1+q^2+p_{\perp}^2)^{1/2}}.
\end{align}
Here $f(q,p_{\bot } )$ is a Fermi-Dirac distribution,  
\begin{equation}
f(q,p_{\bot } )=\frac{1}{1+\exp((\epsilon-\mu)/T)},
\end{equation}
where  $\mu$ is the chemical potential and $T$ is the temperature (we normalize both $\mu$ and $T$ with the electron mass). Non-degenerate initial distributions are still possible with this choice, and can be obtained by picking a large but negative chemical potential. 
 
\begin{figure}
    \centering
    \includegraphics[width=\columnwidth]{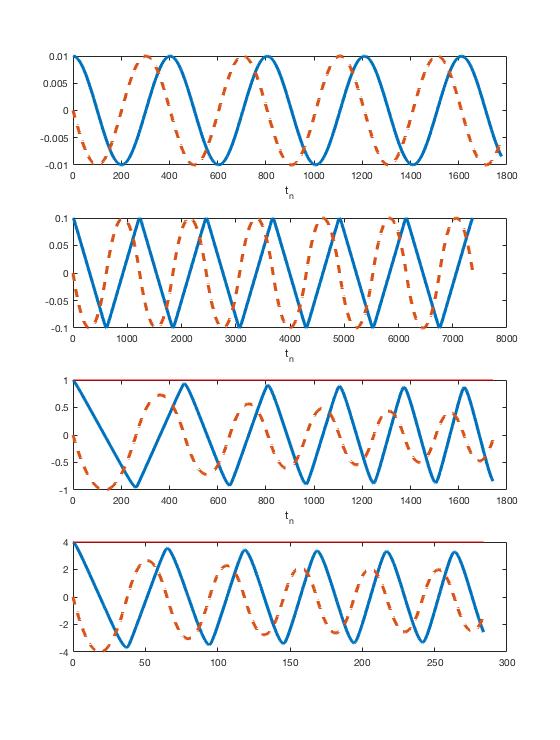}
    \caption{The normalized vector potential $A_{n}$, the dashed lines, and electric field $E$, the curves, are plotted versus the normalized time $t_n=t/\omega_c$ for four different values of initial E-field $E=(0.01,0.1,1,4)$ }
    \label{Figure_1}
\end{figure}
\subsection{Plasma oscillation dynamics}
 In Fig. 1 the electric field and vector potential are plotted for different values of $E_0$, $\mu$, and $T$. To be able to show the evolution of $A$ in the same plot as $E$, we have normalized the vector potential once more, displaying $A_n=E(t=0)A(t)/A_{\rm peak}$, where $A_{\rm peak}$ is the peak absolute value of $A(t)$. Even in the upper panel, with a comparatively modest field strength, $E(t=0)=0.01E_{cr}$, $\mu=1$ and $T=0.1$, we have nonlinear relativistic motion with $A_{\rm peak}=1.85$ (essentially once the motion becomes relativistic, $A_{\rm peak}$ corresponds to the peak gamma factor during an oscillation cycle, averaged over all particles). In spite of the relativistic motion, the deviation from linear behavior is not clearly visible in the wave profile, although a careful analysis would show a harmonic content in the wave form. However, changing the initial electric field to $E(t=0)=0.1E_{cr} $ (still with $\mu=1$ and $T=0.1$), the moderately relativistic motion turns into strongly relativistic oscillations (with $A_{\rm peak}=15.9$. The oscillation is still perfectly periodic, as seen in the second panel of Fig. 1, but now there is a clear sawtooth profile of the electric field. This effect comes from the strong relativistic motion, where the gamma factors are much larger than unity for most of the oscillation, except at the turning points. As a result, for most of the oscillations, all particles move close to the speed of light, giving a current that is more of less constant until it changes direction. Given this, the sawtooth profile of the electric field is a direct consequence of Ampère's law. Moreover, due to the strong relativistic motion, the effective (nonlinear) plasma frequency is much lower, due to the high gamma factors giving the electrons a large effective mass. 
   \begin{figure}
    \centering
    \includegraphics[width=9cm, height=12cm]{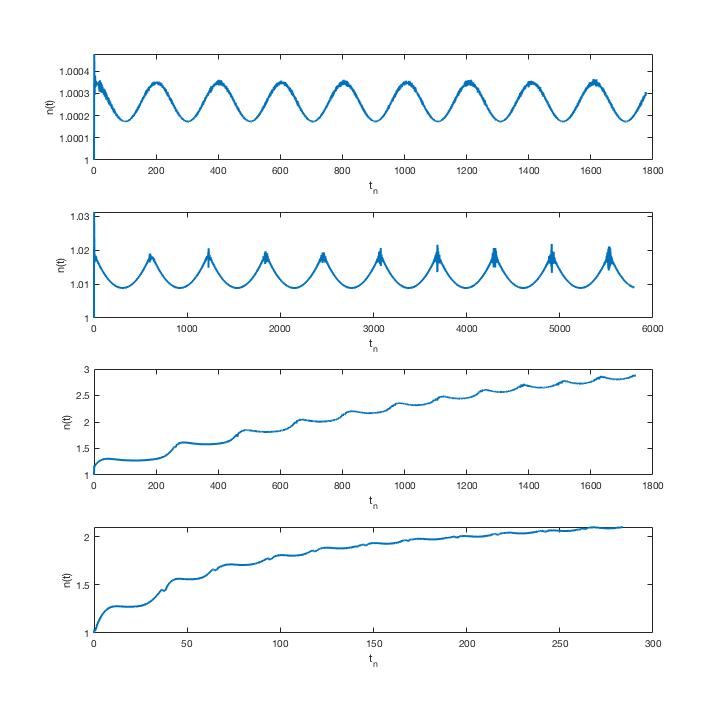}
    \caption{The total particle number density $n(t)$ is plotted over time for four different initial values of the electric field $E(t=0)=(0.01,0.1,1,4)$.  }
    \label{Figure_2}
\end{figure}

 However, the results in the two upper panels do not deviate much from a computation based on the relativistic Vlasov equation (in the standard classical version of Plasma physics, see e.g. Ref. \cite{plasma quantum kinetic review} for a comparison of quantum and classical versions of the Vlasov equation). To see quantum relativistic physics, we need to approach the Schwinger critical field, i.e. let $E(t=0)\sim 1$. In the third panel, we have used $E(t=0)=1$, $\mu=1.5$ and $T=0.2$. Here we still see a sawtooth profile for the electric field as expected, since $A_{\rm peak}=67.2$. However, we now also have pronounced decrease of the vector potential for each oscillation. While the electric field also decreases, we note that $A$ decreases more rapidly than $E$. With $E=-\partial A/\partial t$ the more rapid decrease of $A$ is consistent with an increase in the plasma frequency. While the energy loss of the electric field due to pair-production is clearly seen, the increase in plasma frequency (due to the increased number density) is even more pronounced. The latter effect is seen both in the relation between $E$ and $A$, and by observing the gradual change in time period between successive peaks.  
 
 Finally, in the fourth panel, by picking $E(t=0)=4$, $\mu=4$ and $T=0.2$, corresponding to $A_{\rm peak}=39.7$ we have displayed the dynamics of the plasma oscillation in a regime well beyond the Schwinger critical field. The curve might look surprisingly similar to the third panel with $E(1=0)=1$, with the same type of energy loss and decrease in frequency. However, from the temporal scale we see that here the oscillation frequency is much higher, due to the higher value of the electron number density.  Hence the energy loss due to pair-production is indeed more rapid with a higher initial electric field, as expected. 
 
 That the energy-loss and frequency increase is the result of electron-positron pair production can be  confirmed by investigating the evolution of the total particle (electron and positron) number density $n$. In terms of the normalized $\chi$-variables, $n$ can be expressed as \cite{Perpendicular}

\begin{equation}
\label{Number_pair}
    n= \frac{1}{(2\pi)^3}\int\frac{1}{\epsilon} \left[\epsilon_{\perp}{\chi_2}+ (q-A)\chi_1 \right]d^2p 
\end{equation}
Plotting the quantity $n(t)/n(t=0)$ in \cref{Figure_2}, for the same initial data used in the four panels of Fig 1, we see that there is as a gradual increase of particles in panels 3 and 4, consistent with the energy loss and increase of plasma frequency seen in the corresponding panels of Fig 1. 

At the same time, it is also of some interest to note the fast but small-scale oscillations of $n$ seen in panels 1 and 2. At the peak of the number density in the oscillation cycle, the particle data is noisy, due to zitterbewegung  creating small scales in momentum space. The increase in particle density, however, is reversible, and thus there is no long-term gradual particle increase. The common belief has been that such fluctuations in particle number, as seen in panels 1 and 2 during an oscillation cycle, is artificial. According to this view, while the electromagnetic field is varying dynamically, the particle number is not really well-defined. Instead, to make an unambiguous computation of $n$, the time-asymptotic limit must be considered, when the electromagnetic field decay to zero. However, this view has recently been challenged in an interesting paper, see Ref. \cite{Anton-new}. There it is shown that a high pair-count in the middle of a continuous cycle actually can be realized as measurable pairs, provided the field can be turned off rapidly. On the other hand, in our case, with a self-consistent plasma oscillation field, there is no way to make this happen. Thus, in what follows, when we speak of pair-production, we will be referring to the accumulative process seen in panels 3 and 4, rather than the reversible small-amplitude oscillations seen in panels 1 and 2.

Interestingly, the damping of the electric field is relatively modest even when the pair creation is rather high, as seen when comparing the third and fourth panels of \cref{Figure_1} and \cref{Figure_2}. The reason is that the energy loss by creating low-energy pairs is of the order of two rest mass energies, whereas the characteristic energy of the plasma particles is much higher, larger by a factor $\gamma_{\rm char}$. Here the characteristic value of the $\gamma$-factor typically is $\sim 35$ (with the corresponding peak value $\gamma_{\rm peak}\sim 70$).
As a result significant creation of low-energy pairs can take place without heavy damping.

Furthermore, in regard to \cref{Figure_2}, while the relative increase in number density is somewhat higher in panel 3 as compared to panel 4, the pair-production is considerably more rapid for the higher field, as can be seen when comparing the temporal scales. Finally, we note that there is a saturation of the number density in panels 3 and 4. For a lower field intensity, the decreasing field strength is a partial explanation, but for a higher field strength, the saturation of the damping is mostly due to the created pairs filling the available low-energy states (Pauli blocking). In this context, see e.g. Refs. \cite{Bloch,Tanji:2008ku} where the difference between scalar and spinor QED is studied, starting from vacuum.

\begin{figure}
    \centering
    \includegraphics[width=\columnwidth]{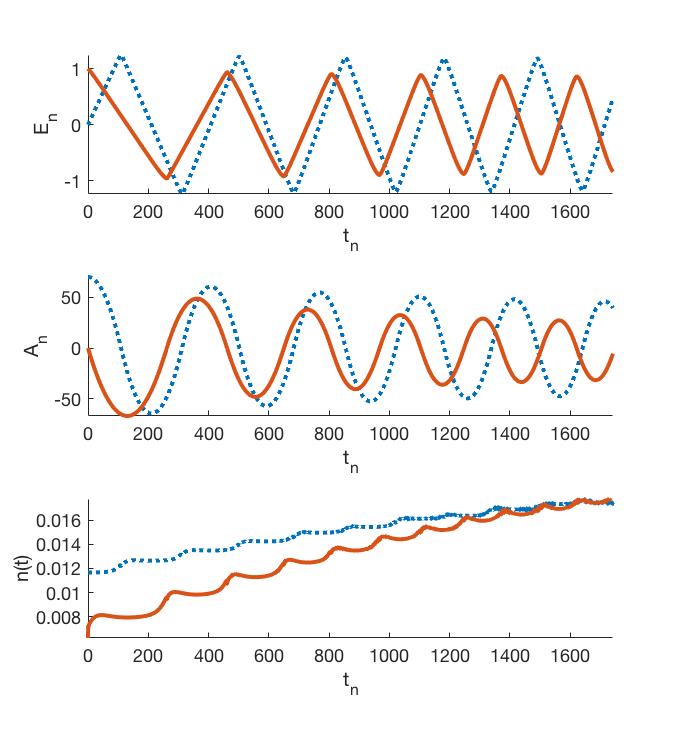}
    \caption{A plot of the electric field (upper panel), vector potential (middle panel), and number density (lower panel) for a plasma with an initial current. The dotted line (blue) correspond to the initial parameters $E=0$, $\mu=3$, $T=0.2$, and $p_d=70$. The solid line (brown) correspond to initial values $E=1$, $\mu=1.5$, $T=0.2$, and $p_d=0$. }
    \label{drift}
\end{figure}

For the case of a plasma initially present, there is no need to have the initial energy stored in the electric field. Starting instead with a sufficiently strong plasma current, that is with a kinetic energy density of the order $\sim \varepsilon_0 E_{\rm cr}^2/2$, the kinetic energy will be converted into an electric field, and the pair production will set in. In \cref{drift} the evolution of the electric field and the vector potential is plotted for the initial conditions of $p_d=70 $, $E=0$, $\mu=3$, and $T=0.2$, where $p_d$ is the kinetic drift momentum of the electrons relative to the stationary ions. For comparison, the results are compared with similar oscillations starting from no initial drift but with $E=1$ $\mu=1.5$ and $T=0.2$. Once the kinetic energy has been converted into an electric field of the order of the Schwinger critical field, the pair creation dynamics shows similar type of damping and frequency up-shift as seen in the plots starting from an electric field and a zero current. A few differences between the two sets of curves, although relatively minor ones, can be noted. The most obvious is that the number of pairs created is a bit larger in the case with no initial drift (lower panel). As a result, the damping of the plasma oscillation is somewhat less pronounced (upper and middle panel). This is not a result of much physical significance, though, as the exact details of the evolution depend on the combination of all initial parameters. It was necessary to pick a higher value of $\mu$ and hence a higher plasma density, for the case of a drifting plasma, in order to get peak electric field of the same order. Nevertheless,
the overall conclusion is that the qualitative dynamics is independent of the initial phase chosen for the plasma oscillation.

\subsection{Analytical estimates}
In order to gain some qualitative understanding, let us make certain analytical
estimates of the pair-creation process. While we are interested in the
dynamics with an initial plasma present, let us first consider the pair-creation
rate starting from vacuum. As long as the available particle states are free
to be filled, the pair-creation rate $dn/dt$, is given by (see e.g. \cite{Schwinger}) 
\begin{equation}
\frac{dn}{dt}=\frac{1}{4\pi ^{3}c}\left( \frac{eE}{\hbar }\right)
^{2}\sum_{n=1}^{\infty }\frac{1}{n^{2}}\exp \left( -\frac{n\pi E_{cr}}{E}%
\right) \label{pair-rate}
\end{equation}%
For electrical fields well above the critical field, in normalized units, we
thus have
\begin{equation}
\frac{dn}{dt}\sim \frac{E^{2}}{4\pi ^{3}}
\end{equation}%
which should be applicable as an order of magnitude estimate also for $%
E\gtrapprox E_{cr}$. Since the created particles will move close to the
speed of light soon after creation, the current generated from the pairs can
be written as $j\sim \int_{0}^{t}(dn/dt^{\prime })dt^{\prime }$, in which
case the electric field energy will be completely converted to particle
energy during a time $T\sim $ $2\pi /(\alpha E_{0})^{1/2}$ (consistent with the results presented in Refs \cite{Gelis-Tanji-review,Gavrilov:2008fv,Gavrilov:2007hq}), which constitute a
quarter of an oscillation cycle, implying an oscillation frequency of the
order
\begin{equation}
\omega \sim \frac{1}{4}(\alpha E_{0})^{1/2}\text{,}
\end{equation}%
where $E_{0\text{ }}$ is the initial electric field. After the electrostatic
energy has been converted to particle energy, the peak $\gamma -$factor for
the created particles $\gamma _{p}$ will be of the order $\gamma _{p}\sim
E_{0}/\omega $, and the average $\gamma -$factor will roughly be   
\begin{equation}
\left\langle \gamma \right\rangle \sim \frac{\gamma _{p}}{2}\sim \frac{E_{0}%
}{2\omega }\sim 2\left( \frac{E_{0}}{\alpha }\right) ^{1/2}.\label{24}
\end{equation}%
Moreover, the number of particles that have been created imply a plasma
frequency $\omega _{p}^{2}$ (with $\omega _{p}^{2}=ne^{2}/\varepsilon _{0}m$
by definition, not accounting for the relativistic frequency lowering of the
wave frequency) 
\begin{equation}
\omega _{p}^{2}\sim \omega ^{2}\left\langle \gamma \right\rangle \sim \frac{1%
}{8}\alpha ^{1/2}E_{0}^{3/2}
\end{equation}%
With a plasma initially present, if the value of $\omega _{p}^{2}$ at $t=0$
is much larger than $\sim \alpha ^{1/2}E_{0}^{3/2}/8$, naturally the
influence on the plasma dynamics due to the Schwinger mechanism will be
comparatively modest. Nevertheless, the accumulated effect during a large
number of periods may still be noticeable. 

Next we assume, for the sake of
the argument, that the initial plasma density is of the same order as the
Schwinger induced contribution after a wave period. This case is considered
in the two lower panels of Figs 1 and 2, and as a consequence we see that
the plasma density roughly increases by a factor 2 during the first
oscillation period. While this implies a substantial increase in both the
plasma density and the wave frequency, interestingly we see that the wave
damping of the plasma oscillation still is relatively modest, as shown by
the electric field evolution. To understand this, for simplicity we
concentrate on the specific case with $E(t=0)$ $\sim 1$, in which case the
pairs will be generated with initial energies $\simeq 2m$. Thus for each
created pair, the plasma oscillation energy will decrease by an amount of the order $%
\sim 2m$. However, subsequent acceleration of the pairs just converts the
electrostatic plasma oscillation energy into kinetic plasma oscillation
energy. As a result, no further wave damping occurs due to acceleration of
the created pairs. Thus we can estimate the relative drop in plasma
oscillation energy during a wave period as 
\begin{equation}
\frac{\delta W}{W}\sim \frac{\delta n_{0}}{\left\langle \gamma \right\rangle
\left( n_{0}+\delta n_{0}\right) }\sim \frac{\delta n_{0}}{\left(
n_{0}+\delta n_{0}\right) }\frac{\alpha ^{1/2}}{2E^{1/2}}.
\end{equation}%
This shows that even with a substantial increase in particle number density
and wave energy, i.e. with $\delta n_{0}\sim n_{0}\,$, the decrease in wave
energy during an oscillation is only a few percent for $E$ of the order
unity. Moreover, for $E>1\,$, the damping of the plasma oscillation
amplitude will not increase with the initial electric field, but rather the
opposite. Naturally, for $E<1$, the exponential dependence included in \cref{pair-rate} will suppress the influence of the Schwinger mechanism.  

As seen in Figs 1 and 2, and confirmed by the present analysis,
the change in wave frequency is a much more pronounced effect than the
plasma oscillation damping (i.e., the energy loss as captured by the decreasing peak electric field), which explains why the vector potential $A$
falls off more rapidly than the electric field. As an order of magnitude
estimate,  the relative increase in frequency during a wave period is 
\begin{equation}
\frac{\delta \omega }{\omega }\sim \frac{\delta n_{0}}{\left( n_{0}+\delta
n_{0}\right) }\sim \frac{E^{1/2}}{\alpha ^{1/2}}\frac{\delta W}{W}. \label{estimate}
\end{equation}%
The above scaling can be deduced using $\omega^2\sim\omega_p^2/\left\langle \gamma \right\rangle$ together with $\delta\omega/\omega\sim -\delta\left\langle \gamma \right\rangle/\left\langle \gamma \right\rangle$, as follows from \cref{24}, given that the electric field amplitude is conserved to zero order in $\alpha$. 

\begin{figure}
    \centering
    \includegraphics[width=\columnwidth]{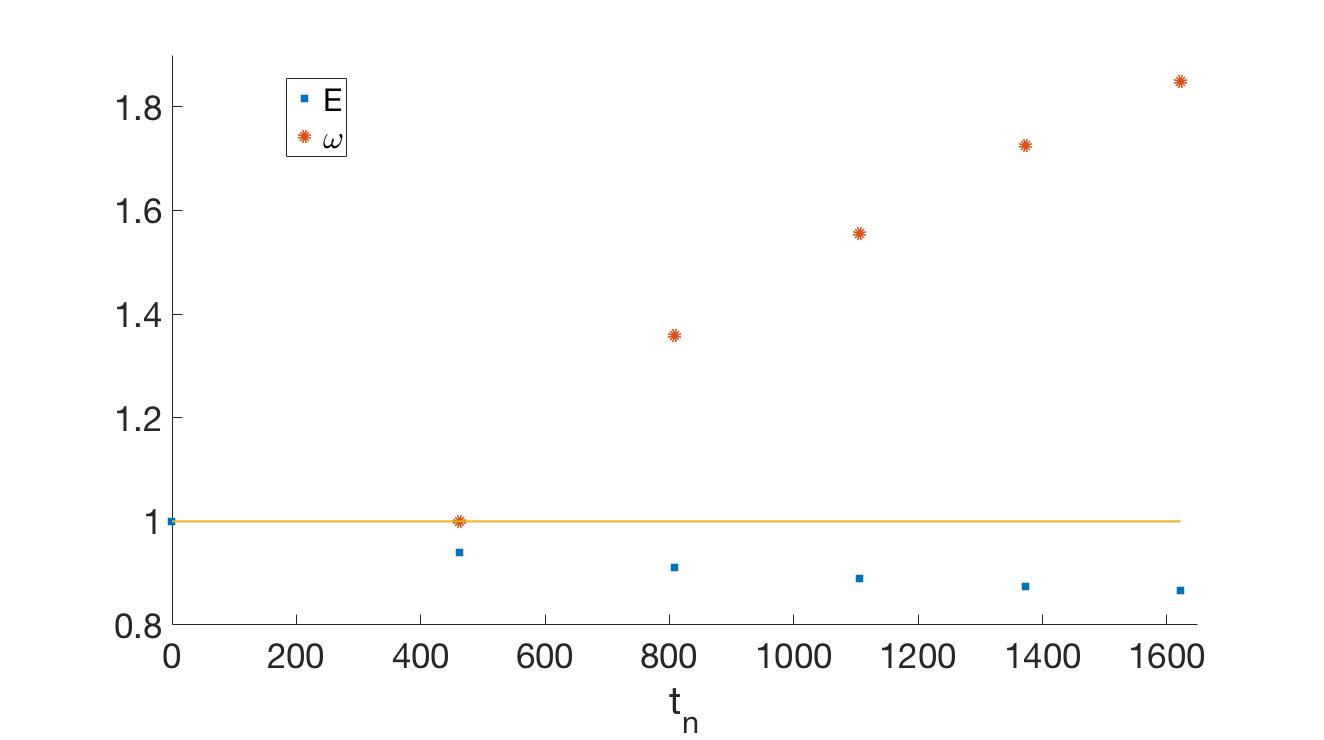}
    \caption{A plot of the electric field damping (squares) and frequency increase (dots) for $E(t=0)=1$, $\mu=1.5$ and $T=0.2$, where a line equal to unity is drawn for comparison.  The electric field dots are evaluated at successive peaks, and the frequencies are computed as $\omega=2\pi/T$,  with the time period $T$ computed for successive peak values of the electric field. The frequencies are normalized against the frequency for the first oscillation period (with $T$=462)}
    \label{frequencyanddamping}
\end{figure}

That the frequency upshift is a more pronounced effect than the damping (larger by a factor $E^{1/2}\alpha ^{-1/2}$ as indicated by \cref{estimate}) is also illustrated in \cref{frequencyanddamping}, where the pair-creation damping and frequency up-shift has been evaluated for the same parameters as in panel 3 of Fig. 1. We note that the electric field is reduced by roughly $10\%$ after five oscillation periods, whereas the frequency has been nearly doubled during the same time.   Although in the next subsection we will study wave damping in more detail, we note that the scaling relation
\cref{estimate} can be used to provide a rough estimate for the corresponding frequency upshift.

\subsection{Plasma oscillation damping}
\begin{figure}
    \centering
    \includegraphics[width=\columnwidth]{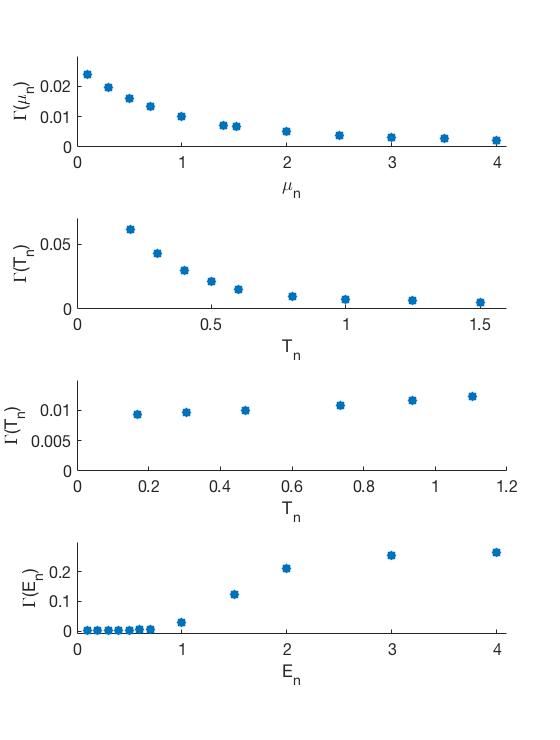}
    \caption{The damping of the electric field $\Gamma$ is plotted as a function of three different parameters. In the upper panel we plot $\Gamma$ versus $\mu$ for $E=1$ and $T=1$ .For the second panel, we plot $\Gamma$ versus the temperature $T$ for $E=1$ and $\mu=1.5$. For the third panel, we vary $T$ but compensate with varying $\mu$ to keep the number density fixed.
    Finally, in the fourth panel, we plot $\Gamma$ versus initial electric field for $E=4$ for $T=0.4$ and $\mu=1.5$.}
    \label{Damping_gamma}
\end{figure}

 The damping of plasma oscillations and the increase in plasma frequency, due to electron-positron pair-production, is the main focus of our work. While the estimates provided in the previous sub-section will give a rough idea of the scaling, important effects such as Pauli-blocking preventing pairs from being created is not accounted for. Accordingly, we will study how the pace of the damping  scales with the plasma parameters (the temperature and the chemical potential) as well as the initial electric field-strength. We define the relative one-period electric field damping $\Gamma$ as
 \begin{equation}
     \Gamma =\frac{\abs{E(T)-E(t=0)}}{E(t=0)},
 \end{equation}
where $T$ is the period time of the first oscillation. In \cref{Damping_gamma} the damping $\Gamma$ is plotted for different values of $E$, $\mu$ and $T$. In the upper panel of \cref{Damping_gamma}, we use the initial electric field $E(t=0)=1$, a temperature $T=1$ and a varying $\mu$. We see that there is a clear decay in the damping rate with increasing chemical potential. The main reason for the decay in damping with $\mu$ is due to the increase in plasma frequency, as the pair-creation rate grows continually during a plasma oscillation cycle. In the second panel, where $\mu$ is fixed but the temperature is varied, there is a similar effect, as the particle density grows with $T$ for a fixed $\mu$. In principle, we could expect a reversal of this mechanism for low temperature, as there would be no free low-energy states to create pairs for a fully degenerate system. While this mechanism certainly is present in principle, we cannot see the effect for this value of $\mu$, as the particle density is too low to block a sufficient volume in momentum space. 

In order to investigate the effect of blocked low-energy states, and to separate the dependence on temperature from that on the initial number density, in the third panel we vary the temperature, but compensate with varying $\mu$ such as to keep the initial number density fixed. Contrary to the second panel in  \cref{Damping_gamma}, we now see an increase in $\Gamma$ with temperature. As argued above, this can be explained as a partial pair-production blocking for the colder system, with less empty low-energy states available where pairs can be produced. 

Finally, in the fourth panel of \cref{Damping_gamma}, for $\mu=1.5$ and $T=0.4$, we see a rapid increase in $\Gamma$ as the initial electric field approaches and pass the critical field. At first one may be a bit surprised to see the saturation in damping occurring for $E(t=0)\sim 3-4$. However, it should be noted that the energy loss rate after a cycle is $\propto E^2\Gamma$, such that even if $\Gamma$ is more or less saturated, the pair-production rate continue to increase with the field $\propto E^2$.

\begin{figure}
    \centering
    \includegraphics[width=\columnwidth] {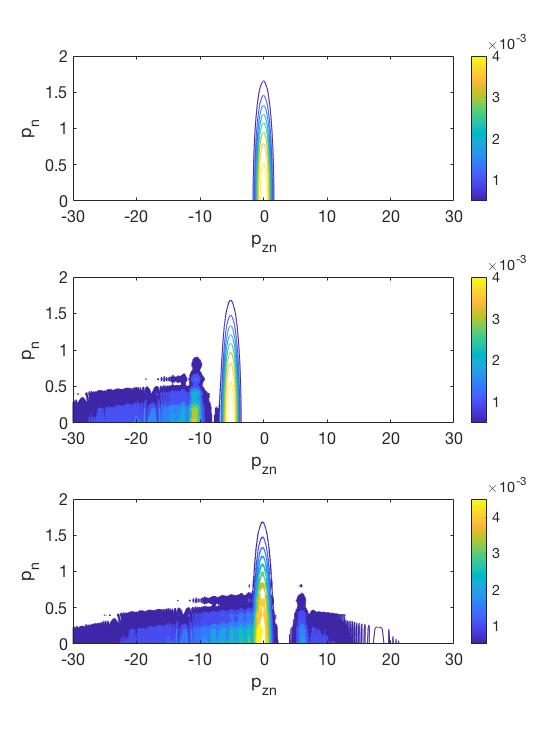}
    \caption{Color mappings of the momentum distribution of the total particle density $n(p_z,p_{\bot})$ at three different times $t=(0, \frac{T}{4},T/2)$.}
    \label{Contour}
\end{figure}

Next, there is a question where in the momentum space the new pairs are produced. In \cref{Contour} contour curves over the integrand in \cref{Number_pair} is shown, $n(p_{\perp},p_z)$, where we have switched back to kinetic momentum $p_z$ rather than canonical momentum $q$. The upper panel shows the contour curves for the initial Fermi-Dirac distribution. A quarter of a plasma period later, we can clearly distinguish the initial particle distribution. As can be seen, the initial particles has been shifted a distance $\approx A(t)$ in the $p_z$-direction, but with a more or less conserved shape. However, in addition to the original particles, a contribution from electron-positron pairs have been added, accelerated by the fields to have a much larger parallel momentum than perpendicular momentum. For a more detailed discussion about the parallel and perpendicular momentum distribution of the produced pairs see Ref.  \cite{Krajewska}. Note that after a quarter period, the produced pairs have larger negative $p_z$ momentum than the original particles. In the lower panel, after a half cycle, the symmetry has been restored, as the produced pairs are located on both sides (in parallel momentum space) of the original particles. As the pair-production process continues, eventually it will be difficult to separate the original particles from the newly produced ones, as the produced particles will outnumber the original ones. 

Before we end the discussion of the pair-creation dynamics in the presence of plasma, let us point out that there are interesting papers of relevance in this context by Refs. \cite{Bulanov,Marklund-etal2006}.  Firstly, in Ref. \cite{Bulanov}, the authors consider an initial plasma in the context of Schwinger pair-production. However, rather than studying an electrostatic field, they compute the damping of an electromagnetic wave due to the energy loss through the Schwinger mechanism.  However, in contrast to our study, that paper considered a cold plasma with a delta function for the initial momentum distribution, which limits the potential applicability to rather modest plasma densities (with a small Fermi momentum) and temperatures.  

Finally, in Ref. \cite{Marklund-etal2006}, production of pairs by electrostatic waves in plasma was studied using a relativistic quantum fluid model assuming that the electric field varies slowly in comparison with the Compton frequency. While the overall pair production rate used in hydrodynamics based on the Schwinger theory (proportional to $\exp(-\pi E_{cr}/(E))$) is consistent with the DHW-formalism, the kinetic details (e.g. where in momentum space the pairs are created) are important for the plasma oscillation properties. More specifically, many features seen in our treatment such e.g. Pauli blocking, cannot be covered in a hydrodynamic approach. Nevertheless, the more modest computational demands makes hydrodynamical studies a good complement to kinetic treatments, which is illustrated by the fact that the spatial influence also could be covered in Ref. \cite{Marklund-etal2006}.    
\subsection{The initial vacuum regime}
Up to now we have  considered pair-creation dynamics in the presence of a plasma. By contrast, previous works of a similar nature, e.g. Refs. \cite{Bloch,Hebenstreit:2013qxa,Buyens:2016hhu,Kasper:2014uaa,Kasper:2016mzj,Kasper:2015cca,Gelis:2013oca,Tanji:2008ku}, have been focusing on pair-creation dynamics when starting from a vacuum state. In this case, naturally one cannot begin by postulating a plasma oscillation with a fixed amplitude as initial condition, as there is no initial plasma. Instead one has to use an external electric field starting up the process, either in the form of a time-dependent Sauter-pulse \cite{Sauter} or by imposing a constant external field at $t=0$.  In spite of these differences, once the plasma is created, many of the features seen in the present paper when considering a plasma are the same as those seen in previous works starting from vacuum. In particular, the plasma oscillation amplitude tend to diminish gradually and the plasma oscillation frequency increases due to pairs being gradually created \cite{Cooper-89,Bloch,Kluger}. 

However, starting from vacuum is numerically more challenging than studying the physics of a high-density plasma, due to the very large parallel momenta induced by a field close to the Schwinger limit, unless the initial plasma density and thereby plasma frequency is high. As a result, most previous numerical works have simplified the numerical schemes in various ways. For example, some works (e.g. Refs. \cite{Kluger,Hebenstreit:2013qxa}) limited the treatment to include no perpendicular momentum dependence in the calculation. While several of the previous works (e.g. Ref. \cite{Bloch}) account also for the dependence on perpendicular momenta, importantly, most previous calculations starting from vacuum and including the back-reaction through Ampère's law have simplified the numerics by using a fine structure constant that is much larger than the actual value. As it turns out, this choice limits the peak parallel momentum rather drastically. This is not entirely unproblematic when a complex dynamical situation is studied. While the basic physical mechanisms does not change when using a magnified $\alpha$, the relative importance of the dynamical mechanisms will not stay the same. 

In order to compare with previous works, particularly Ref. \cite{Bloch}, we have started the dynamics with an external Sauter-pulse of the form 
\begin{align}
 A_{ex}(t)&=A_0 \bigg[1 +\tanh{\Big(\frac{t-t_{0}}{\tau } \Big)}\bigg]\\
 E_{ex}(t)&=-\frac{A_0}{\tau \cosh^2{\Big(\frac{t-t_{0}}{\tau } \Big)}}
\end{align}
and used the actual value of the fine structure constant $\alpha=0.007297$ with $\eta =\alpha/(2\pi)^2$. Starting the numerical computation at $t=0$, we let $t_0=-2$ in all our runs, to include a short build-up phase of the Sauter-pulse. For the above case, the tendency is that almost all pair-creation happens in the very first initial stage, much shorter than a plasma period, before any plasma oscillations can be seen. 

\begin{figure}
    \centering
    \includegraphics[width=\columnwidth]{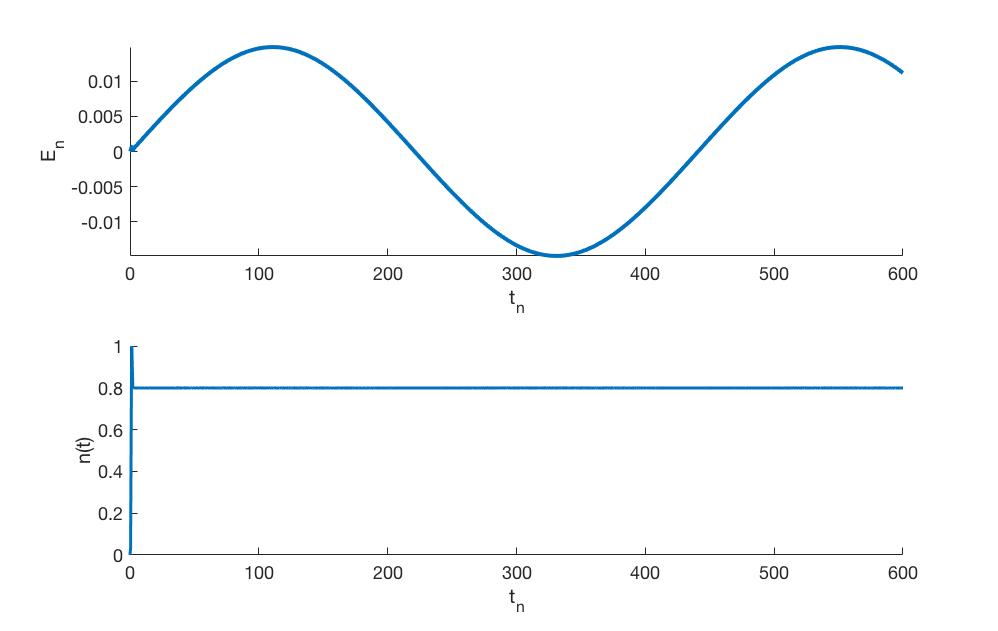}
    \caption{
      Plot of the induced electric field (i.e. without including the external field of the initial Sauter pulse) (upper panel), and the induced number density (lower panel). The parameter values for the Sauter pulse used here are $A_0=1$ and $\tau=0.5$ }
    \label{A0=1}
\end{figure}

This is illustrated in
\cref{A0=1}
where we have used the parameters $A_0=1$ and $\tau=0.5$.  As the density immediately (compared to the plasma oscillation time scale) reaches its peak value, no visible pair creation is done by the self-consistent field that is responsible for the plasma oscillation seen in the two upper panels. This is perhaps not surprising, as the electric field strength of the self-consistent field (upper panel) is below the critical field by orders of magnitude, atlhough the peak field of the Sauter pulse is fairly strong (of the order of the critical field).

\begin{figure}
    \centering
    \includegraphics[width=\columnwidth]{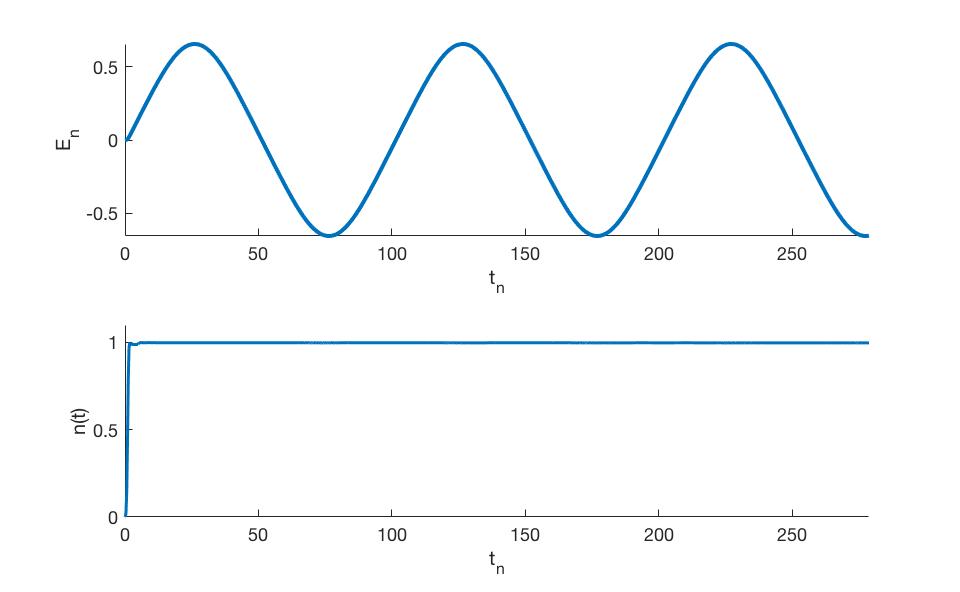}
    \caption{Plot of the induced electric field (i.e. without including the external field of the initial Sauter pulse) (upper panel), and the induced number density (lower panel). The parameter values for the Sauter pulse used here are $A_0=10$ and $\tau=0.5$} 
    \label{Blochnopair}
\end{figure}

Next, we consider a much  stronger Sauter pulse with $A_0=10$ and $\tau=0.5$. The result is shown in \cref{Blochnopair}. As can be seen in the upper panel, the self-consistent field is much stronger (peak electric field around $0.6 E_{cr}$), but the qualitative dynamics is still close to the previous case. In particular, the peak electric field of the Sauter pulse is far from recovered, and the extra pair-creation that takes place after the initial phase is more or less negligible, as seen from the constant number density in the lower panel.

\begin{figure}
    \centering
    \includegraphics[width=\columnwidth]{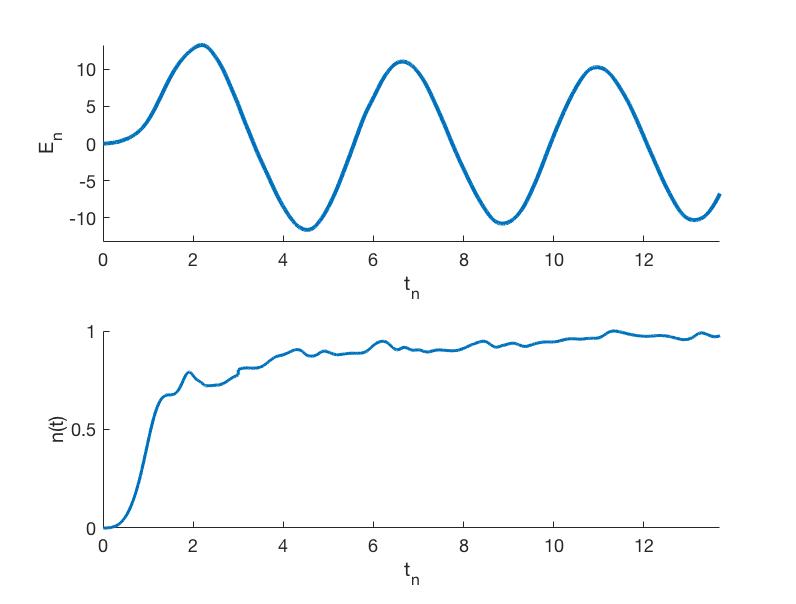}
    \caption{
      Plot of the induced electric field (i.e. without including the external field of the initial Sauter pulse) (upper panel)) and the induced number density (lower panel). The parameter values for the Sauter pulse used here are $A_0=10$ and $\tau=0.5$ and the enhanced value $\alpha=4$ for the fine structure constant.}
    \label{Enhancedalpha}
\end{figure}

As our next step, in order to illustrate the influence of a modified value of $\alpha$ on the dynamics, we consider an enhanced value of the fine structure constant. For this purpose, we match the parameters of Ref. \cite{Bloch}, also accounting for the charge renormalization (with a cut-off value $\Lambda=25)$), the magnified value of the coefficient in Ampère's law become $\eta=0.26$ (as compared to $\eta=\alpha/(2\pi^2)\approx 0.3697\times 10^{-3}$).  In \cref{Enhancedalpha} the electric field and pair density is shown for a Sauter pulse with  initial parameters $A_0=40$ and $\tau=0.5$. With a peak field of $E\sim 15$, and a similar oscillation frequency and decay in the upper panel as previously found by Ref. \cite{Bloch}, there is close resemblance to results computed for a magnified value of $\alpha$. In the lower panel of \cref{Enhancedalpha}, we see the evolution of the number of pairs. While much of the particles are created by the initial Sauter-pulse, there is subsequent significant pair-creation due to the plasma field (generated by the back-reaction), accompanied by further damping of the electric field. However, as is apparent from a comparison with \cref{Blochnopair}, the dynamics depend rather sensitively on the value of $\alpha$. 



\begin{figure}
    \centering
    \includegraphics[width=\columnwidth]{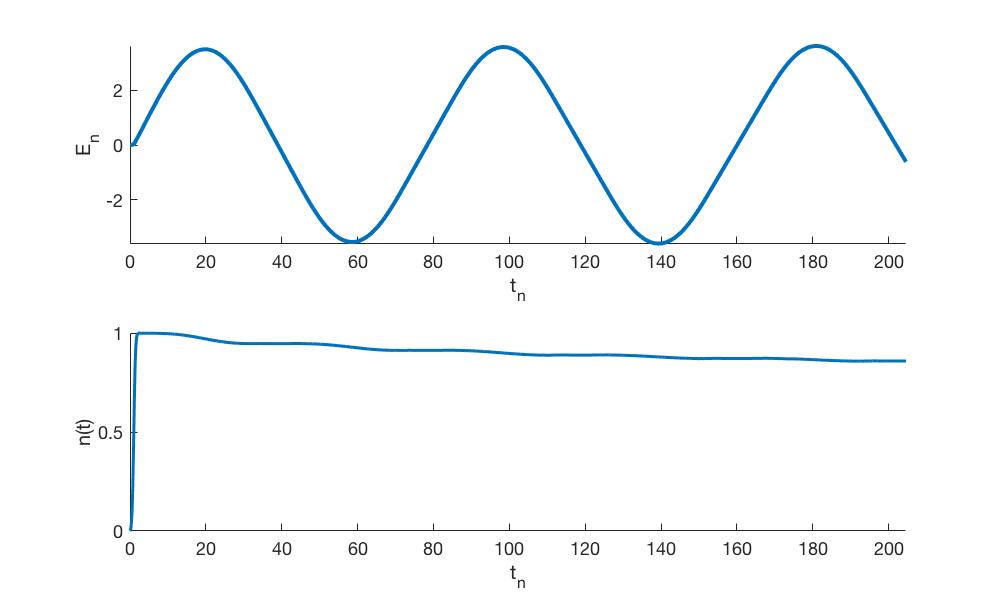}
    \caption{
      Plot of the induced electric field (i.e. without including the external field of the initial Sauter pulse) (upper panel), induced vector potential (middle panel), and number density of the produced pairs (lower panel). The parameter values for the Sauter pulse used here are $A_0=40$ and $\tau=0.5$ }
    \label{Blochpair}
\end{figure}

From now on, we return to the real value of the fine-structure constant. In \cref{Blochpair}, the evolution is shown for the initial parameters $A_0=40$ and $\tau=0.5$. Just as for the weaker pulses shown in \cref{A0=1} and \cref{Blochnopair}, still there is not much pair-creation after the initial stage. Interestingly, though, rather the reverse process can be seen, as illustrated in the plot of the number density in the lower panel of \cref{Blochpair}. Here we see that during the later stage of the evolution, the pair density decreases with roughly $14 \%$ compared to the peak value. Accordingly we have a collective dynamic where annihilation is more prominent than pair-creation, even though both processes can take place simultaneously in different parts of momentum space. Interestingly, although the effect is too small to be clearly visible in the upper panel, we can note a slight {\it increase} in the electric field amplitude. To illustrate this feature the peak amplitude has been explicitly written out in the  upper panel of \cref{Blochpair}.

Usually one expects annihilation processes to generate gamma rays. In this context, the main point is that the collective annihilation mechanism gives energy to the electric field of the plasma oscillation, rather
than producing hard photons. Still, in principle there will also be collisional annihilation, that can produce gamma rays, an effect not covered by the present study. The corresponding cross-section is small, however, and thus neglecting collisional annihilation is a good approximation.


\begin{figure}
    \centering
    \includegraphics[width=\columnwidth]{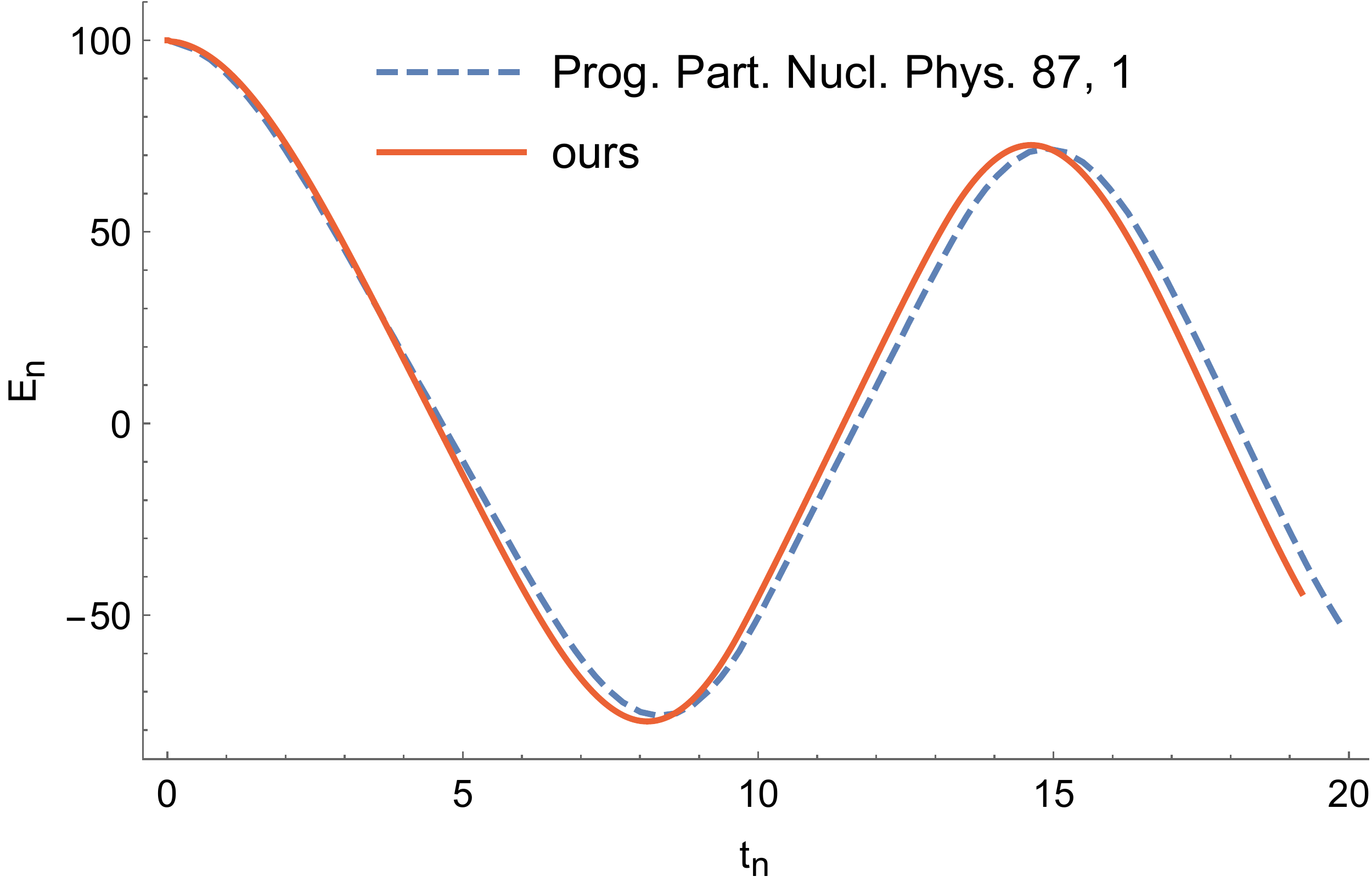}
    \caption{The electric field profile starting from a an initial value $E(t=0)=100 E_{cr}$ using the actual value of $\alpha$. The result is compared with the curve computed in Ref. \cite{Gelis-Tanji-review}, using the same parameters. The data from \cite{Gelis-Tanji-review} has been extracted using the online tool WebPlotDigitizer \cite{digitizer}}
    \label{GT-comparison}
\end{figure}
While the use of a magnified fine structure constant has been rather common in the research literature when starting from vacuum, some works (e.g. Refs. \cite{Kasper:2014uaa, Gelis-Tanji-review} have used the actual value. Specifically, we now compare our results with those of \cite{Gelis-Tanji-review}. Here the authors started from a homogeneous electric field as initial condition (rather than a Sauter pulse). This is a numerically challenging problem, due to large relativistic factors (requiring a large volume in momentum space) combined with the distinct separation of time scales, with the Compton scale being much shorter than the plasma oscillation time scale. However, for very strong initial fields, the problem turns out to be less demanding computationally, as the separation in time scales is not equally pronounced. To confirm the agreement with previous works in this regime, we have used a very strong initial field $E(t=0)=100 E_{cr}$, together with the actual value of $\alpha$, and compared the electric field profile with the curve shown in Ref. \cite{Gelis-Tanji-review} using the same parameters, see \cref{GT-comparison}. While there is a small but visible difference between the two curves, the deviation is not larger than should be expected with regard to the numerical precision.  

As the dynamics when the back-reaction is prominent (for a discussion of when back-reaction can be neglected, see e.g. Refs. \cite{Gavrilov:2008fv,Gavrilov:2007hq}) tend to depend on the actual value of the fine-structure constant, it would be of interest to systematically re-consider many of the studies made using a magnified value of $\alpha$. Since this tend to be computationally demanding, and our main focus has been on the regime with an initial plasma, a more complete study of the initial vacuum regime is outside the scope of the present paper. 

\section{Competing mechanisms}
\label{Competing mechanisms}
In this section, we will consider potential mechanisms that have been omitted, but may compete with the processes considered up to now. In the overall dynamics, the main effects of pair production is a frequency increase of the plasma oscillations, together with wave damping. The present DHW-model is an accurate description of the mean-field physics, but processes converting energy from coherent degrees of freedom to non-coherent degrees of freedom is not covered. For high plasma densities, we note that in principle it is possible that certain incoherent processes outside the mean field description can be suppressed due to many-particle effects. For a concrete example illustrating the interplay between, long-scale collective phenomena, and short-scale incoherent processes, see e.g. Fig 1 in Ref. \cite{Gonoskov:2014mda}.

Specifically, single particle radiation  emitted by accelerated particles can be suppressed when the generated frequencies $\omega_e$ fulfill $N\equiv nc^3/\omega_e^3\gg1$, indicating that a large number of electrons $N$ simultaneously interact with the same photon quanta. A precise estimate of the  characteristic frequency $\omega_e$ would require substantial research, considering that the emission depend on the details of the particle distribution and may vary a lot during the oscillation period. Here we just note that soft emission (with $\omega_e\ll \omega_c$), generally will have $N\gg1$, in which case collective interaction captured in the mean field description will suppress single-particle effects. However, due to field strengths of the order the critical field, and a spread in perpendicular momentum not much smaller than $m$, a significant part of the emitted spectrum is likely to have frequencies $\omega_e$ of the order $\omega_c$ or possibly higher, implying $N<1$, in which case single particle emission gives an extra energy loss mechanism for the plasma oscillations not captured by the DHW-model. Accordingly, in what follows, we will assume that single-particle emission is present, and estimate the energy loss of the plasma oscillations based on this.  

Next we note that the single-particle photon emission implies
an effective friction-like force, where the standard expression is referred to
as the Landau-Lifshitz force \cite{Landau_Lif}. This force scales as $\gamma ^{2}$,
and tend to be important when the field is strong enough to induce large
particle energies. However, to a large degree this effect is suppressed in a
1D electrostatic geometry, when the large gamma factor only comes from the
motion parallel to the field. However, also in this case there is a finite
contribution that survives, due to the particle spread in momentum. Using the classical Landau-Lifshitz expression \cite{Landau_Lif} for the radiation reaction force $\mathbf{F}_{rad}$, which reduces to 
\begin{equation}
    \mathbf{F}_{rad}=\frac{2e^3 \epsilon}{3m^2}
    \bigg[ \Big(\frac{\partial E}{\partial t} + 
    \frac{eE^2p_z}{\epsilon^2}\Big)\mathbf{\hat{z}}
    -\frac{eE^2\epsilon_{\bot}^2}{m^2\epsilon^2}\mathbf{p}
    \bigg]
\label{rad-force}
\end{equation}
in the given field geometry, 
we can estimate the energy density
loss due to the radiation field induced by the particle self-interaction during a
plasma period. Dropping the term proportional to the time-derivative of the electric field in \cref{rad-force}(as this contribution is considerably smaller in the strong field regime of interest here), the energy loss rate $d\Delta W/dt$ of the plasma wave energy is given by~
\cite{footnote}

\begin{equation}
\frac{d}{dt}\Delta W=\big\langle\mathbf{F}_{rad}\cdot \mathbf{v}\big\rangle\approx \frac{2}{3}\omega_p^2r_{0}E^{2}\langle\frac{p_{\bot
}^{2}}{m^2}\rangle \label{loss-rate}
\end{equation}
Here $\langle...\rangle$ denotes averaging over the particle distribution, $r_{0}$ is the classical electron radius and $\omega_p$ is the non-relativistic plasma frequency. It can be noted that $\omega_p=\sqrt{n_0e^2/\epsilon_0 m}$ is related to the wave oscillation frequency (as seen e.g. in Fig. 1), according to $\omega^2\sim \omega_p^2/\gamma$, where $n_0$ is the number density in the laboratory frame, and $\gamma$ is the characteristic relativistic factor of the particles averaged over a wave-period. Using \cref{loss-rate}, the relative energy loss $\Delta W_P/W$ during a wave period can be estimated as
\begin{equation}
 \frac{\Delta W_P}{W}\sim r_{0}\frac{\omega _{p}^2}{\omega}
 \label{period-loss}
\end{equation}
Here we have used $\int (E^2(t)dt/E^2(t=0))\sim 3/\omega$, and $\langle p_{\perp}^2/m^2\rangle \sim 1/2$ for the plasma parameters typically used. 
Using the Lorentz-force equation $dp_z/dt=eE$ we can estimate $\gamma\sim p_z/m\sim eE/m\omega$ and hence
\begin{equation}
\frac{\Delta W_P}{W}\sim\alpha\frac{E}{E_{cr}} \;.   
\end{equation}
Thus for $E$ of the order of the critical field, the energy loss per wave period scales as $\alpha$ rather than $\sqrt{\alpha}$ as for the Schwinger mechanism. While this suggests that radiation reaction gives a smaller effect, comparing with \cref{Damping_gamma} we see that for the actual value of $\alpha$, we cannot be certain. 

On the other hand, we note that an estimate based on the classical expression \cref{rad-force} tend to overestimate the loss rate as compared to a more accurate calculation based on QED, see e.g. \cite{RMP-new-review}. This is captured in the so called Gaunt-factor $G$, which describes the quantum reduction in the energy loss rate. 
This is usually used in cases with large perpendicular momenta (for which a general field appears as a crossed field), while in our case we have large longitudinal momentum instead. Nevertheless, Fig.~14 of Ref. \cite{RMP-new-review}, where the standard $G(\chi_e)$ is plotted, suggests that the quantum corrected result significantly reduces the classical estimate for $\chi_e:=\sqrt{-(F^{\mu\nu}p_\nu)^2}/(mE_{cr})=E p_{\perp}/m E_{cr}\sim 0.5-1$. 

Naturally, for electric field strengths well below the critical field, the exponential suppression of the pair-creation rate  will make radiation losses dominant. For further results concerning field depletion due to radiation losses, see e.g. Ref. \cite{Seipt}. 

To conclude, we expect that for fields of the order of the critical field, the inclusion of radiation reaction would not significantly modify the plasma oscillation dynamics, even though the wave damping due to pair-creation is not necessarily stronger than that due to radiation losses (compare e.g. \cref{Figure_1}). For the case when both mechanisms are small, however, to leading order the plasma oscillation energy loss can be computed perturbatively as the sum over two separate mechanisms while disregarding the other. Importantly, the frequency up-shift due to the increasing number density, as induced by pair-creation, is a much more pronounced effect than the wave damping (compare \cref{estimate}), and this has no correspondence due to radiation reaction. Nevertheless, although radiation reaction does not change the plasma number density, there will still be a frequency increase due to the decreasing (average) gamma factors, as induced by the energy loss, although that effect is one order higher in $\alpha$, as compared to the frequency increase due to the Schwinger mechanism.        

Finally, it should be noted that there is also a possibility that mechanisms such as Breit-Wheeler pair production could affect the dynamics. In this case, similar dynamics as studied here can be expected, i.e. a frequency up-shift due to the produced pairs, see e.g. Ref \cite{Qu:2020gpy}). However, a rigorous analysis of such contributions are outside the scope of the present paper.

\section{Conclusion}
\label{Conclusion}

The main purpose with this study has been to study self-consistent plasma dynamics for ultra-strong field strengths. It is seen that for fields strengths of the order $E_{cr}$ or larger, electron-positron pair-production leads to a pronounced increase in the plasma oscillation frequency, and also a damping in the amplitude. The scaling of the damping with the plasma parameters have been investigated numerically. In particular, for a fixed chemical potential $\mu$, pair-production is suppressed for higher temperatures. This is because the mechanism depends sensitively on the plasma density, which is increased with higher temperature (for a fixed $\mu$). However, if the particle number density is kept fixed instead, there is a slight increase in the damping with temperature. This is because more low energy states will be available for pair-creation. 

For numerical convenience, we have picked fairly high plasma densities in our study, since the code can be executed on a PC in this case. However, from the point of view of laboratory experiments, the case with $\mu$ and $T$ much less than unity is of most interest. While we have not probed this regime here, the scaling behavior displayed in \cref{Damping_gamma} still applies. Nevertheless, extended simulations (run on a parallel computer) must be made for detailed quantitative predictions in the regime where both $\mu\ll 1$ and $T\ll 1$ applies.  

For field strength well beyond $E_{cr}$, the dynamics is  similar to the case $E\sim E_{cr}$, but the processes is considerably more rapid. Moreover, starting with an electron-positron plasma as initial condition, the process can evolve in the opposite direction. That is, instead of Schwinger pair-production, we can have collective Schwinger annihilation, leading to an increase in the oscillation amplitude. While this process is much faster than collisional annihilation, it still competes with pair production taking place in the part of momentum space that is not filled with electrons and positrons. Thus, as far as we have seen, pair-annihilation is typically not a dominant mechanism.

In conclusion, the DHW-formalism is very well suited for studies of collective strong field plasma dynamics. Future problems of interest based on this approach include extensions to non-homogeneous systems, electromagnetic fields \cite{Bulanov,Mueller2016}, and magnetized plasmas.    

%






\begin{thebibliography}{99} 

\bibitem{review 1}
A. Di Piazza, C. M\"uller, K. Z. Hatsagortsyan, and C. H. Keitel,
Rev. Mod. Phys.,
 {\bf 84},1177,
 2012.


\bibitem{recent review}
A Fedotov, A Ilderton, F Karbstein, B King, D Seipt, H Taya, and G Torgrimsson, arXiv:2203.00019 (2022).

\bibitem{ELI}
http://www.extreme-light-infrastructure.eu/.
\bibitem{Schwinger}
J. Schwinger, Phys. Rev. \textbf{82}, 664 (1951).
\bibitem{plasma mirror}
F. Quéré and H. Vincenti, HPLSE   {\bf 9},6 (2021).

\bibitem{Gies}
F. Hebenstreit, R. Alkofer, and H. Gies. Phys. Rev. D \textbf{82}, 105026 (2010).

\bibitem{Gies2}
F. Hebenstreit, R. Alkofer, and H. Gies, Phys. Rev. Lett. \textbf{107}, 180403 (2011).

\bibitem{Aleksandrov:2019ddt} I.~A.~Aleksandrov and C.~Kohlf\"urst, 
 Phys. Rev. D \textbf{101}, 096009 (2020).

 \bibitem{Kohlfurst:2019mag}
C.~Kohlf\"urst,
Phys. Rev. D \textbf{101}, 096003 (2020).

\bibitem{optimal pulse}
A. Gonoskov,I Gonoskov, C. Harvey, A. Ilderton,  A. Kim, M. Marklund, G. Mourou, and A. Sergeev, Phys Rev. Lett, {\bf 111}, 060404, (2013).

\bibitem{Cooper-89}
F. Cooper and E. Mottola, Phys. Rev. D \textbf{40}, 456 (1989).

\bibitem{Kluger}
Y. Kluger, J. M. Eisenberg, B. Svetitsky, F. Cooper, and E. Mottola
Phys. Rev. D \textbf{45}, 4659 (1992).

\bibitem{Ruffini}
R. Ruffini, L. Vitagliano, and S.S. Xue. Phys. Letters B 559.1-2 (2003).

\bibitem{Bloch}
J. C. R. Bloch, V. A. Mizerny, A. V. Prozorkevich, C. D. Roberts, S. M. Schmidt, S. A. Smolyansky, and D. V. Vinnik
Phys. Rev. D \textbf{60}, 116011 (1999).

\bibitem{Hebenstreit:2013qxa}
F.~Hebenstreit, J.~Berges and D.~Gelfand,
Phys. Rev. D \textbf{87}, no.10, 105006 (2013)

\bibitem{Kampfer}
 A Otto et al 2019 Plasma Phys. Control. Fusion \textbf{61} 074002 (2019).
 
\bibitem{Gold:2020qzr}
G.~Gold, D.~A.~Mcgady, S.~P.~Patil and V.~Vardanyan,
JHEP \textbf{10}, 072 (2021)
[arXiv:2012.15824 [hep-th]].

\bibitem{Buyens:2016hhu}
B.~Buyens, J.~Haegeman, F.~Hebenstreit, F.~Verstraete and K.~Van Acoleyen,
Phys. Rev. D \textbf{96}, no.11, 114501 (2017)

\bibitem{Kasper:2014uaa}
V.~Kasper, F.~Hebenstreit and J.~Berges,
Phys. Rev. D \textbf{90}, no.2, 025016 (2014)

\bibitem{Kasper:2016mzj}
V.~Kasper, F.~Hebenstreit, F.~Jendrzejewski, M.~K.~Oberthaler and J.~Berges,
New J. Phys. \textbf{19}, no.2, 023030 (2017)

\bibitem{Kasper:2015cca}
V.~Kasper, F.~Hebenstreit, M.~Oberthaler and J.~Berges,
Phys. Lett. B \textbf{760}, 742-746 (2016)

\bibitem{Gelis:2013oca}
F.~Gelis and N.~Tanji,
Phys. Rev. D \textbf{87}, no.12, 125035 (2013)

\bibitem{Gelis-Tanji-review}
F.~Gelis and N.~Tanji,
Progress in Particle and Nuclear Physics, \textbf{87}, no.12, 1-49 (2016)
\bibitem{Tanji:2008ku}
N.~Tanji,
Annals Phys. \textbf{324}, 1691-1736 (2009)
\bibitem{Sauter}
F. Sauter, Zeit. Phys. \textbf{69}, 742 (1931).
\bibitem{Birula}
I. Bialynicki-Birula, P. Gornicki, and J. Rafelski, Phys. Rev. D \textbf{44}, 1825 (1991).
\bibitem{Vasak-87}
D. Vasak, M. Gyulassy and H. T. Elze, Ann. Phys. \textbf{173}, 462 (1987).
\bibitem{Sheng}
X. L. Sheng, R. H. Fang, Q. Wang, and D. H. Rischke, Phys. Rev. D  \textbf{99}, 056004 (2019).

\bibitem{Estat-note}
 We will refer to the given field as electrostatic, since there is no magnetic field. However, aiming for an electrostatic scalar potential in the homogeneous limit is not possible, as the scalar potential would then be infinite. Sometimes the field used here is seen as the dipole approximation of an electromagnetic field. The possibility to use opposing terminology for the same physical set-up appears because the electrostatic Langmuir wave and the electromagnetic wave coincides in the limit of infinite wavelength.  
\bibitem{Perpendicular}
H. Al-Naseri, J. Zamanian and G. Brodin, Phys. Rev. E
\textbf{104}, 015207 (2021).
\bibitem{plasma quantum kinetic review}
G. Brodin and J. Zamanian,  Rev. Mod. Plasma Phys. {\bf 6}, 4 (2022). 

\bibitem{linear} 
H. Al-Naseri and G. Brodin , Phys. Plasmas {\bf 29}, 042106 (2022) 


\bibitem{Anton-new} 
A. Ilderton
Phys. Rev. D {\bf 105}, 016021 (2022).

\bibitem{Schwinger}
J. Schwinger,
Phys. Rev. {\bf 82}, 664 (1951). 

\bibitem{Gonoskov:2014mda}
A.~Gonoskov, S.~Bastrakov, E.~Efimenko, A.~Ilderton, M.~Marklund, I.~Meyerov, A.~Muraviev, A.~Sergeev, I.~Surmin and E.~Wallin,
Phys. Rev. E \textbf{92}, no.2, 023305 (2015).

\bibitem{Krajewska}
K. Krajewska and J. Z. Kaminski, Phys. Rev. A \textbf{100}, 012104 (2019).
\bibitem{Bulanov}
S. S. Bulanov , A. M. Fedotov, and F. Pegoraro. Phys. Rev. E 71.1 016404 (2005).

\bibitem{Mueller2016}
N. Mueller, F. Hebenstreit, and J. Berges
Phys. Rev. Lett. {\bf 117}, 061601 (2016.) 

\bibitem{Marklund-etal2006}
M. Marklund, B. Eliasson, P. K. Shukla, L. Stenflo, M. E. Dieckmann and M. Parviainen, JETP Letters, {\bf 83}, 313 (2006). 
\bibitem{Gavrilov:2008fv}
S.~P.~Gavrilov and D.~M.~Gitman,
Phys. Rev. Lett. \textbf{101}, 130403 (2008)


\bibitem{Gavrilov:2007hq}
S.~P.~Gavrilov and D.~M.~Gitman,
Phys. Rev. D \textbf{78}, 045017 (2008)

\bibitem{digitizer}
Ankit Rohatgi, WebPlotDigitizer, see https://apps.automeris.io/wpd/




\bibitem{Landau_Lif}
L. D. Landau and E. M. Lifshitz, The Classical Theory of Fields
(Pergamon, Oxford, 1975).


\bibitem{RMP-new-review}
A. Gonoskov, T. G. Blackburn, M. Marklund, and S. S. Bulanov, Rev. Mod. Phys., in press (2022);
ArXiv:2107.02161. 

\bibitem{Seipt} D.~Seipt, T.~Heinzl, M.~Marklund and S.~S.~Bulanov, Phys. Rev. Lett. \textbf{118}, no.15, 154803 (2017).










\bibitem{Qu:2020gpy}
K.~Qu, S.~Meuren and N.~J.~Fisch,
Phys. Rev. Lett. \textbf{127}, no.9, 095001 (2021)





\bibitem{footnote}
Another way of obtaining this is from the Larmor formula, $P(p)=-(2/3)e^2\dot{p}_\mu^2$, by approximating $P(p_{RR})\approx P(p_{Lo})$, where $p_{Lo}^\mu$ is the solution to the Lorentz force equation without RR, $\dot{p}^\mu=eF^{\mu\nu}p_\nu/m$.
 
\end{thebibliography}
\end{document}